\documentclass[11pt]{article}
\usepackage{moriond,epsfig}

\usepackage{psfrag}
\usepackage{feynmf}

\usepackage{pstcol}

\newrgbcolor{purple}{0.6 0.2 0.4}
\newrgbcolor{dgreen}{0.2 0.4 0.2}
\newrgbcolor{blue}{0.0 0.0 0.7}
\newrgbcolor{red}{0.7 0 0}

\def\be{\begin{equation}}
\def\ee{\end{equation}}
\def\bea{\begin{eqnarray}}
\def\eea{\end{eqnarray}}

\newcommand{\gtsim}{{\,\stackrel{>}{\sim}\,}}

\newcommand{\bbms}{$B_s\!-\!\ov{B}{}_s\,$\ mixing}
\newcommand{\bbmd}{$B_d\!-\!\ov{B}{}_d\,$\ mixing}
\newcommand{\bbb}{$B_s\!-\!\ov{B}{}_s\,$}
\newcommand{\lt}{\left}
\newcommand{\rt}{\right}

\newcommand{\ov}[1]{\overline{#1}}

\newcommand{\cL}{{\cal L}}
\newcommand{\cO}{{\cal O}}
\newcommand{\tg}{{\tilde g}}

\newcommand{\mtpole}{{m_t^{\mbox{\rm \scriptsize pole}}}}
\newcommand{\tq}{{\tilde q}}
\newcommand{\td}{{\tilde d}}

\newcommand{\tl}{{\tilde l}}

\newcommand{\tn}{{\tilde \nu}}
\newcommand{\gev}{\ensuremath{\,\mbox{GeV}}}
\newcommand{\mgut}{{M_{\mbox{\rm \scriptsize GUT}}}}
\newcommand{\mten}{{M_{10}}}
\newcommand{\mpl}{{M_{\mbox{\rm \scriptsize Pl}}}}
\newcommand{\bra}[1]{\langle \, #1 \, | }
\newcommand{\ket}[1]{| \, #1 \, \rangle }

\newcommand{\tchi}{{\tilde \chi}}

\newcommand{\Bslash}{B \!\!\!\! / }

\begin{document}
\rightline{PITHA-05/07}
\vspace*{4cm}
\title{\boldmath \bbb\ MIXING AND LEPTON FLAVOR VIOLATION IN SUSY
\boldmath $SO(10)$~\footnote{Contribution to the XXXXth Rencontres de Moriond ``Electroweak
Interactions and Unified Theories'', La Thuile, Italy, March 5-12, 2005}
}

\author{ S. J{\"A}GER~\footnote{Email address: sjaeger@physik.rwth-aachen.de} }

\address{RWTH Aachen, Institut f\"ur Theoretische Physik E,\\
D-52056 Aachen, Germany}

\maketitle\abstracts{
We present a quantitative analysis of the flavor physics of a SUSY $SO(10)$ model
proposed by Chang, Masiero, and Murayama, linking $b \to s$ transitions to the
large observed atmospheric neutrino mixing. We consider \bbms\ and $\tau \to \mu \gamma$ and
discuss their correlation paying careful attention to their different dependences on model parameters.
Our analysis shows that an order-of-magnitude enhancement of the mass difference with respect
to the Standard Model is possible and can be reconciled with the most stringent bounds on
lepton flavor violation. The situation for various $\Delta B=1$ decays is discussed qualitatively.
}

\section{Introduction}
\label{sec:intro}
Within the Standard Model, all flavor structure (before electroweak
symmetry breaking) resides in the Yukawa coupling matrices. However,
not all of their parameters are physical. For instance,
$$\cL_\mathrm{SM} \supset - \,H \,\bar q_{Li}\, Y^D_{ij}\, d_{Rj} +
         \frac{1}{3} \, g \,\bar d_{Ri} \,\Bslash \,d_{Ri},
\qquad Y^{D} = U_{L}^{D} \hat Y^{D} U_{D} ,
$$
where the hat here and below denotes a diagonal matrix,
and the field redefinition $d_{Ri} \equiv U^\dagger_{Dij}\, d'_{Rj}$
eliminates $U_{D}$ without affecting any other term in the Lagrangian,
leaving it unphysical.

In the MSSM additional flavor structure appears in the soft-breaking
terms, for instance,
$$
\cL_{\mbox{\small soft}} \supset
- \td_R^\dagger m^{2}_{\td_{R}} \td_R,
\qquad  m^{2}_{\td_{R}} = \tilde U \hat m^{2}_{\td_{R}} {\tilde U}^{\dagger} .
$$
As a consequence, the quark-squark-gluino vertex $\bar d_{Ri} \td_{Rj} \tg$
acquires a flavor-changing structure $(U_D\, {\tilde U})_{ij}$.
To avoid the associated large flavor-changing neutral currents (FCNC), it is common to
consider flavor-universal soft-breaking
terms, as in minimal supergravity (mSUGRA). While the
universality in mSUGRA only holds at a high (Planck or GUT) scale,
for the right-handed down-type squark masses the radiative corrections are
small if $\tan \beta$ is not too large, as there are no contributions due to
the large top Yukawa coupling. The same is true of the slepton masses.

Within SUSY GUTs, quarks and leptons are unified into irreducible symmetry
multiplets.
Radiative corrections due to $y_t$ above
the GUT scale then affect full GUT multiplets.
This can generate mass differences
and flavor violation in
the down-type-squark
and the slepton mass matrices,
and the mixing angles contained in $U_D \,{\tilde U}$  can be rendered observable.\cite{BarbieriSUSYGUT}

Furthermore, hadronic and leptonic flavor structures are linked in GUTs.
The presence of large atmospheric and solar mixing angles
in the leptonic sector raises the question whether they can manifest themselves in the hadronic sector.
Chang, Masiero, and Murayama (CMM) have proposed an $SO(10)$ model in which the
large $\nu_\mu$--$\nu_\tau$ mixing angle can affect transitions
between right-handed $b$ and $s$ quarks.\cite{cmm}

\section{Renormalization and particle spectrum in the CMM model}
The $SO(10)$-symmetric superpotential has the form~\cite{cmm}
\be
W_{10} = \frac{1}{2} \mathbf{16}^T Y^U \mathbf{16} \, \mathbf{10}_\mathbf{H} 
  \, + \, 
   \frac{1}{\mpl} \frac{1}{2} \mathbf{16}^T \tilde{Y}^D  \mathbf{16}
     \, \mathbf{10_H^\prime} \mathbf{45_H} 
 + \frac{1}{\mpl} \frac{1}{2}  \mathbf{16}^T Y^M \mathbf{16}  
   \, \mathbf{\ov{16}_H} \mathbf{\ov{16}_H}. \label{w10}
\ee
Here the vector $\mathbf{16}$ in generation space comprises the three
generations of matter, and $Y^U$ is a symmetric
$3\times 3$ matrix containing the large top Yukawa
coupling. The Higgs fields $\mathbf{10_H}$ and $\mathbf{10'_H}$
contain the MSSM Higgses $H_u$ and $H_d$, respectively.
The dimension-5 terms, suppressed by the Planck mass $\mpl$,
involve further Higgs fields  $\mathbf{45_H}$ and $\mathbf{\ov{16}_H}$
acquiring VEVs $v_{45}$ and $v_{\ov{16}}$ of order the scale
$\mten$, where $SO(10)$ is broken to $SU(5)$.
Below the GUT scale $\mgut$, there are the usual four Yukawa matrices of the
MSSM (including right-handed neutrinos), with
$Y^D\propto \tilde Y^D \frac{v_{45}}{\mpl} \approx {Y^E}^T$
and $Y^U \approx Y^\nu$.
Note also that generically $\tan\beta = \cO(\mten/\mpl)$.
The last term in
Eq.~\ref{w10} generates small seesaw neutrino masses.
Flavor symmetries are then invoked which
render $Y^U$ and $Y^M$ simultaneously diagonal. This defines the
$(U)$-basis. In this basis the remaining Yukawa matrix
$Y^D \approx {Y^E}^T$, giving mass to down-type
quarks and charged leptons, contains all flavor mixing:
\be             \label{eq:yukd}
        Y^D = V_{\rm CKM}^*\,\left(\matrix{y_d \cr & y_s \cr && y_b}\right)\,
                U_{\rm PMNS}.
\ee
Here $V_{\rm CKM}$ and $U_{\rm PMNS}$
are the quark and leptonic mixing matrices and
certain diagonal phase matrices have been omitted. We drop the subscript
on $U_{\rm PMNS}$ below.

The soft SUSY-breaking terms are assumed universal near the Planck
scale.
The only large Yukawa coupling $y_t$ now renormalizes the $SO(10)$ sfermion
mass matrix, keeping it diagonal in the $(U)$-basis
but suppressing the masses of the third generation.
At the weak scale, the right-handed down-type squark and the left-handed
slepton mass matrices thus take the form
$$
 m^{2,(U)}_{\td_{R}} =
\left( \matrix{ m^2_{\td_{R1}} & 0 & 0 \cr
0 & m^2_{\td_{R1}} & 0 \cr
0 & 0 &
m^2_{\td_{R1}} - \Delta_\td } \right),
\qquad
  m^{2,(U)}_{\tl} =
\left( \matrix{ m^2_{\tl_1} & 0 & 0 \cr
0 & m^2_{\tl_1} & 0 \cr
0 & 0 &
m^2_{\tl_1} - \Delta_\tl } \right) .
$$ 
This nonuniversal structure induces flavor-changing couplings
once the charged-lepton and down-type-quark
Yukawa matrices are diagonalized, as discussed
in the introduction.
From Eq.~\ref{eq:yukd} and the GUT relation $Y^D \approx {Y^E}^T$,
it follows that the PMNS mixing matrix $U$ governs
the coupling of right-handed down-type squarks to gluinos and quarks as well as
that of left-handed sleptons to charginos and neutralinos and leptons.
In particular, the large atmospheric mixing angle will appear in the mixing of
${\tilde b}_R$ and ${\tilde s}_R$.
Given the apparent lack of Yukawa
unification for the first two generations, which due to their smallness are
more sensitive to higher-dimensional operators,\cite{yukunif} it may not be justified to
expect a complete correspondence of leptonic and hadronic mixing angles.
Imposing (approximate) Yukawa unification only for the third
generation is sufficient to obtain
$
        |{U_D}_{23}| \approx |{U_D}_{33}| \approx |U_{\mu 3}|
        \approx |U_{\tau 3}| \approx 1/\sqrt{2},
$
suggesting large FCNC involving second-to-third-generation
transitions. On the other hand, $|{U_D}_{13}| \approx |U_{e3}|$
is unknown and consistent with zero, so (for nearly degenerate
first two generations of sparticles) no large effects occur in
$1 \to 2$ and $1 \to 3$ transitions. (For instance, the \bbmd\ 
phase remains $2 \beta$,
which ensures consistency with the standard CKM fit.)
For this reason, the
phenomenological analysis below is restricted to $2 \to 3$ processes.

Besides $\tan\beta$, we choose as low-energy input parameters
the approximately universal first-generation squark mass $m_\tq$,
the gluino mass $m_\tg$, and a down-squark $A$-parameter
$a_d \equiv A^{(U)}_{D11}/y_d$.
We use the MSSM and GUT RGEs to relate them
to the Planck-scale parameters $m_0$, $a_0$,
and $m_{1/2}$. The low-energy quantities
$\Delta_\td$ and $\Delta_\tl$ then are calculable functions of our
low-energy inputs, with the biggest contributions coming from the running
above $\mten$. As a further input we have the (supersymmetric)
$\mu$-parameter, which we take as a free low-energy input.

A renormalization-group study of the top Yukawa coupling
shows that predictivity of the model
requires $y_t$ to remain below a certain value, which translates
to a lower bound on $\tan \beta \gtsim 3$, depending on the top mass
and (mildly) on supersymmetric parameters.\cite{jntalks}
The largest effects occur at this ``critical'' value. We take
$\mtpole=178 \gev$ and set $\tan\beta$ to its critical value in the phenomenological analysis below.

\section{\bbms}
\begin{figure}
\vspace*{-1mm}
\begin{center}
\psfragscanon
\psfrag{mgl}{}
\psfrag{msq}{$m_\tq$}
\psfrag{adbymsq}[Bl][Bl][1][0]{$a_{d}/m_\tq$}
\parbox{60mm}{\vskip-4mm \hskip-5mm
        \resizebox{82mm}{!}{\includegraphics{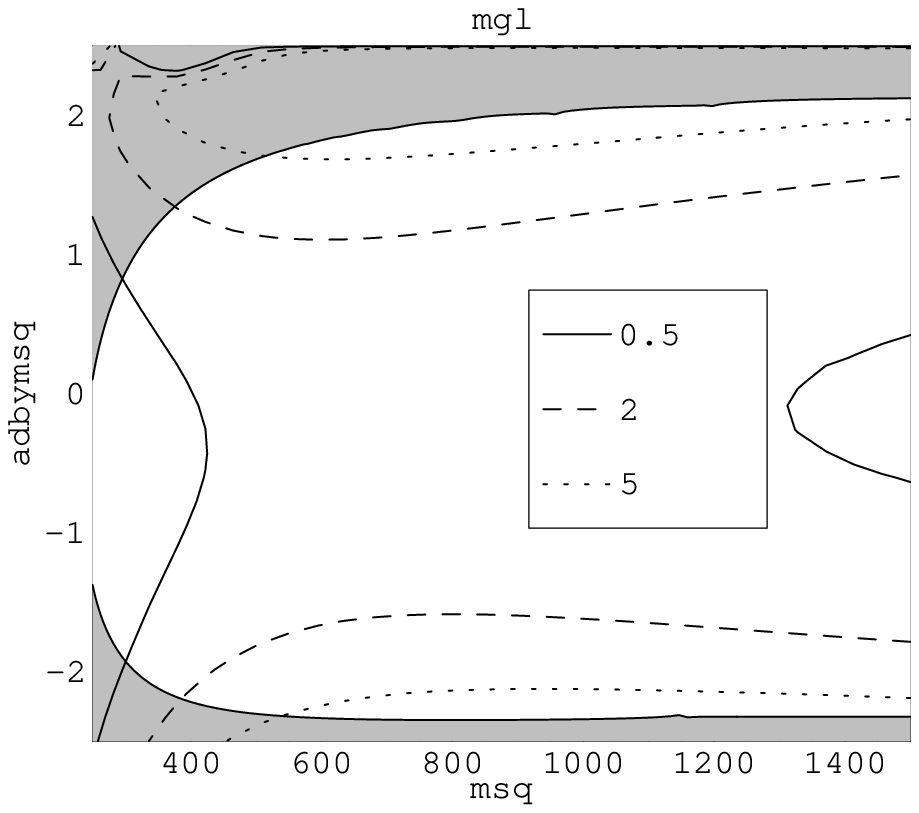}}
}
\hspace{20mm}
\parbox{75mm}{\vskip-4mm
  \psfragscanon
  \psfrag{Re2M12}{\parbox{18mm}{\Large $\mbox{Re}\left(2 M_{12}\right)$ \\
                                \hspace*{2mm} $[\mbox{ps}^{-1}]$}}
  \psfrag{Im2M12}{\Large $\mbox{Im}\left(2 M_{12}\right) [\mbox{ps}^{-1}]$}
  \psfrag{mmm150}{\Large $-150$}
  \psfrag{mmm75}{\Large $-75$}
  \psfrag{pp75}{\Large $75$}
  \psfrag{pp150}{\Large $150$}
  \resizebox{70mm}{!}{\includegraphics[]{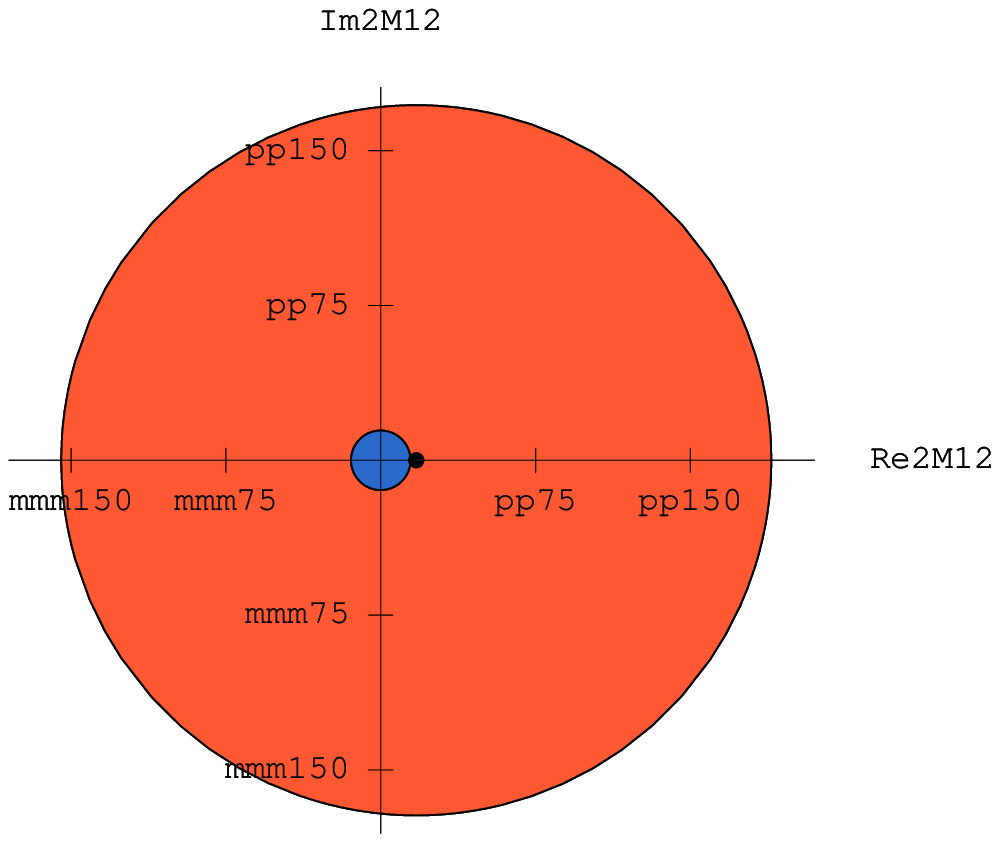}}
  \psline{->}(-1.2,4.5)(-4.05,3.05)
  \rput[bl](-1,4.5){\parbox{20mm}{SM\\ 17.2 ps$^{-1}$}}
}
\caption{{\it Left:} New-physics contribution to \bbms, normalized to
SM value, for $m_{\tg}=250 \gev$.
{\it Right:} Allowed region in the complex $M_{12}$ plane, see text for
explanation.
\label{fig:bb}}
\end{center}
\end{figure}

The dominant new contributions to \bbms\ in the CMM model are $\cO(\alpha_s^2)$
corrections from one-loop box 
diagrams with gluinos and squarks. One obtains the mixing amplitude~\cite{jntalks}
\be
 M_{12} = \frac{G_F^2\, M_W^2 }{32  \pi^2 \, M_{B_s}} \, \lambda_t^2
        \, 
          \lt( C_L + C_R \rt) \bra{B_s} O_L
        \ket{\ov{B}_s} . \label{eq:m12}
\ee
Here $ \bra{B_s} O_L \ket{\ov{B}_s}$
is the hadronic matrix element of the Standard Model effective operator
$O_L = \bar s_L \gamma_{\mu} b_L \; \bar s_L \gamma^{\mu} b_L$ and
$C_L$ is due to Standard Model $W\!-t\!$ exchange. The only new operator generated is
the parity reflection of $O_L$, with the Wilson coefficient
\be
        C_R = 
        \frac{\Lambda_3^2}{\lambda_t^2} 
        \frac{8 \pi^2 \alpha_s^2 (m_{\tilde g}) }{G_F^2 M_W^2 m_{\tilde g}^2 }
        \left[ \frac{\alpha_s(m_{\tilde g})}{\alpha_s(m_b)}\right]^{6/23}
        S^{(\tg)}  \label{eq:CR}
\ee
and the same matrix element as $O_L$.
$\lambda_t= V_{ts}^* V_{tb}$ and
$
        |\Lambda_3| = |U_{\mu 3}| |U_{\tau 3}| \approx \frac{1}{2}
$
denote the relevant flavor-mixing parameters in the Standard Model and the
CMM model, and $S^{(\tg)}$ is a dimensionless function of the
squark and gluino masses. Note the enhancement of $C_R$ due
to the large atmospheric mixing and the large $\alpha_s$.
The neutral $B_s$-meson mass difference is given
by $\Delta M_{B_s} = 2 | M_{12} |$,
while the phase of $M_{12}$ is responsible for mixing-induced CP
violation. Fig.~\ref{fig:bb} (left) shows a contour plot of the modulus
of the new physics contribution normalized to the Standard Model prediction in
the $(m_\tq,a_d)$ plane for $m_\tg=250$~GeV.
The shaded area in the plot is excluded by experimental constraints on the
sparticle spectrum.
We observe that the Standard Model prediction of about $17.2\,\mathrm{ps}^{-1}$
can be exceeded by up to one order of magnitude. The effect decreases rapidly
with increasing gluino mass.
As the phase of $\Lambda_3$ is undetermined,
there is potentially large mixing-induced CP violation in decays such as
$B_s \to \psi \phi$ and $B_s\to\psi \eta^{(\prime)}$. This is also
visible from Fig.~\ref{fig:bb} (right), where the accessible region in
the complex $M_{12}$ plane is shown in light grey (orange) and the region
excluded by the experimental lower bound~\cite{bhadprop} $\Delta M_{B_s}>14.5\,\mathrm{ps}^{-1}$
is shown in dark grey (blue). The SM
prediction, almost real, is also indicated.

\section{$\Delta F=1$ processes: Constraints and predictions}
\subsection{Constraint from $BR(\tau\to\mu\gamma)$}
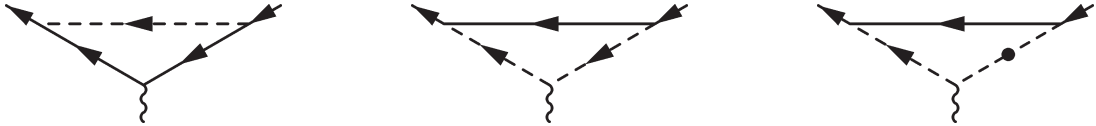
\begin{figure}
\begin{center}
\[
\begin{fmffile}{tmgpengproc}
\parbox{37mm}{
\begin{fmfgraph*}(104,49) \fmfpen{thin}
\fmftop{e1,e2}
\fmfbottom{e3}
\fmflabel{$\mu_L$}{e1} \fmflabel{$\tau_R$}{e2}
\fmflabel{$\gamma$}{e3}
\fmf{fermion}{v1,e1}
\fmf{fermion}{e2,v2}
\fmf{photon}{v3,e3}
\fmf{fermion,label=$\tchi^-_i$,l.side=left,tension=0.3}{v3,v1}
\fmf{fermion,label=$\tchi^-_i$,l.side=left,tension=0.3}{v2,v3}
\fmf{dashes_arrow,tension=0.0,label=$\tn_k$}{v2,v1}
\end{fmfgraph*}}
\hspace{17mm}
\parbox{37mm}{
\begin{fmfgraph*}(104,49) \fmfpen{thin}
\fmftop{e1,e2}
\fmfbottom{e3}
\fmflabel{$\mu_L$}{e1} \fmflabel{$\tau_R$}{e2}
\fmflabel{$\gamma$}{e3}
\fmf{fermion}{v1,e1}
\fmf{fermion}{e2,v2}
\fmf{photon}{v3,e3}
\fmf{fermion,label=$\tchi^0_i$,tension=0.0}{v2,v1}
\fmf{dashes_arrow,tension=0.3,label=$\tl^-_k$,l.side=left}{v3,v1}
\fmf{dashes_arrow,tension=0.3,label=$\tl^-_k$,l.side=left}{v2,v3}
\end{fmfgraph*}}
\hspace{17mm}
\parbox{37mm}{
\begin{fmfgraph*}(104,49) \fmfpen{thin}
\fmftop{e1,e2}
\fmfbottom{e3}
\fmflabel{$\mu_L$}{e1} \fmflabel{$\tau_R$}{e2}
\fmflabel{$\gamma$}{e3}
\fmf{fermion}{v1,e1}
\fmf{fermion}{e2,v2}
\fmf{photon}{v3,e3}
\fmf{fermion,label=$\tchi^0_i$,tension=0.0}{v2,v1}
\fmf{dashes_arrow,tension=0.3,label=$\tl^-_k$,l.side=left}{v3,v1}
\fmf{dashes,tension=0.6,label=${{\tilde \tau}}_R$,l.side=left}{v2,v4}
\fmf{dashes,tension=0.6,label=$\tl^-_k$,l.side=left}{v4,v3}
        \fmfv{decor.shape=circle,decor.filled=1,decor.size=2thick}{v4}
\end{fmfgraph*}}
\end{fmffile}
\]
\end{center}
\vskip2mm
\caption{Diagrams contributing to $\tau \to \mu \gamma$. Approximate mass eigenstates
of three different masses are involved, two for the left-handed sleptons and one right-handed
charged slepton.
\label{fig:taumugdiag}}
\end{figure}

In the leptonic sector a strong constraint comes from the process $\tau\to\mu\gamma$.
The Standard Model contribution is negligible,
and the supersymmetric contributions~\cite{taumugsusy} are due to
the diagrams shown in Fig.~\ref{fig:taumugdiag}.
Note that the amplitude involves a chirality flip, either via the equation of
motion on the external $\tau$ line, through a Yukawa-like coupling
in the vertex, or through an internal LR flip due to left-right mixing
in the slepton mass matrix (only for the neutralino contribution).
In the CMM model, this LR mixing is small and can be treated
in the mass-insertion approximation.
Such an expansion would not be well justified for the LL mixing due to
the large mixing angle present there.
The branching ratio takes the (schematic) form
$$ BR(\tau \to \mu \gamma) \propto |C_7'|^2, \qquad
C_7' = \frac{e^3}{(4\pi)^2 \sin^2 \theta_W}
U_{\tau 3} U_{\mu 3}^*
A \left( m_\tl, m_{\tchi^{0,-}}, \mu \right) .
$$
As an important difference to \bbms, it depends on
the $\mu$-parameter, which is uncorrelated with the squark and slepton masses
in the CMM model. Numerically we find that the chargino-sneutrino diagrams
dominate the amplitude for $|\mu| < 1000 \gev$. Furthermore,
the dominant piece exhibits decoupling as $1/\mu$ for $|\mu| \to \infty$.
\begin{figure}
\begin{center}
\parbox{70mm}{
  \psfragscanon
  \psfrag{mm6}{$10^{-6}$}
  \psfrag{mm7}{$10^{-7}$}
  \psfrag{mm8}{$10^{-8}$}
  \psfrag{mm9}{$10^{-9}$}
  \psfrag{muGeV}{$\mu$[GeV]}
 \resizebox{75mm}{!}{\includegraphics[]{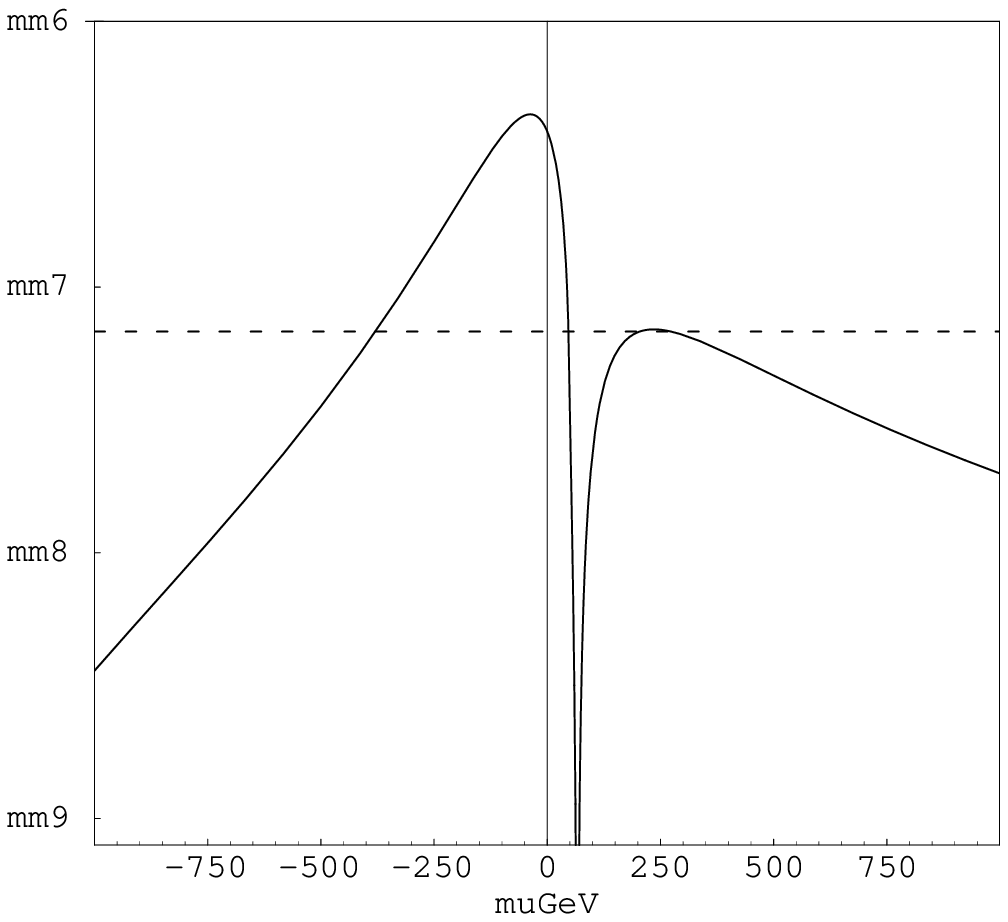}} \\
}
\hspace{7mm}
\parbox{80mm}{\vskip-3mm
\psfragscanon
\psfrag{m6}{$10^{-6}$}
\psfrag{mexp}{$6 \cdot 10^{-7}$}
\psfrag{m6p5}{$10^{-6.5}$}
\psfrag{m7}{$10^{-7}$}
\psfrag{m7p5}{$10^{-7.5}$}
\psfrag{m8}{$10^{-8}$}
\psfrag{m9}{$10^{-9}$}
\psfrag{mgl}{$m_{\tg_3}$}
\psfrag{msq}{\Large $m_\tq$}
\psfrag{adbymsq}[Bl][Bl][1][180]{\Large $a_{d}/m_\tq$}
\resizebox{80mm}{!}{\includegraphics{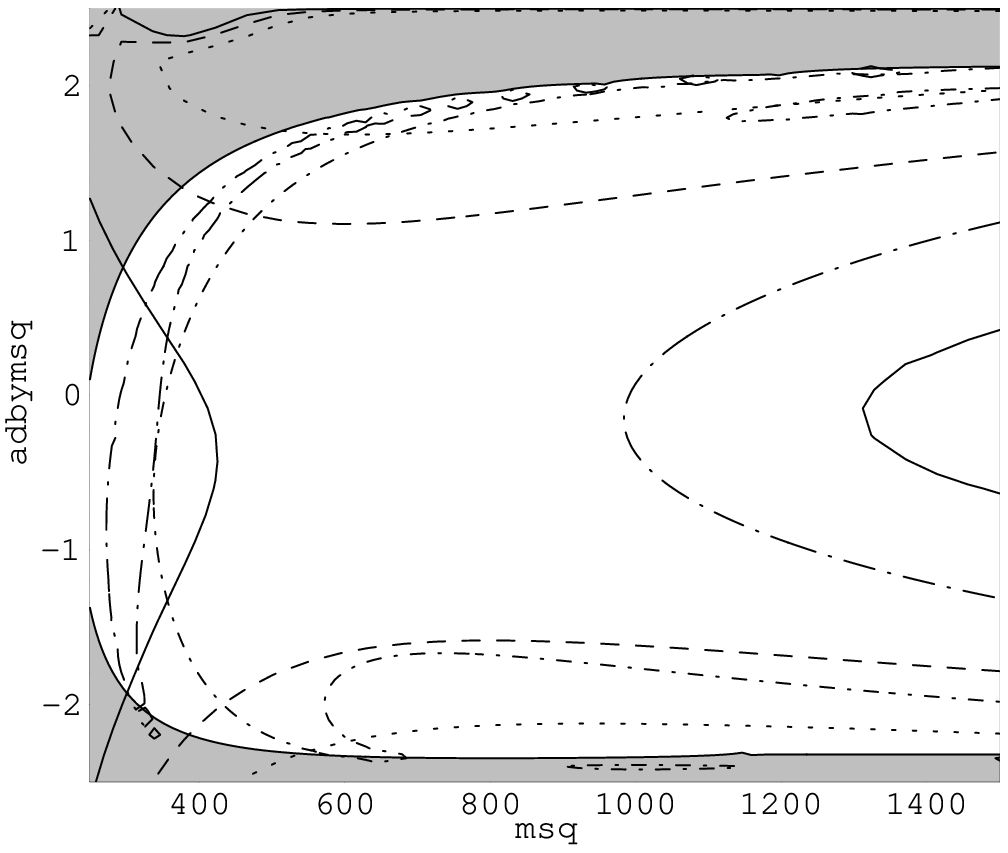}}
}
\end{center}
\caption{{\it Left:} Dependence of $BR(\tau \to \mu \gamma)$ on
$\mu$. The dashed line corresponds to the experimental upper bound.
For parameter values besides $\mu$, see text.
{\it Right:} Same as Fig.~\ref{fig:bb} (left), but including contours where the
experimental limit on $BR(\tau \to \mu \gamma)$ is reached. The
long-dash-dotted line corresponds to $\mu=-300 \gev$,
the short-dash-dotted one to $\mu=-600 \gev.^c$
\label{fig:bbtmg}}
\end{figure}
The effect of this is visible in Fig.~\ref{fig:bbtmg} (left), where
the dependence of the branching ratio (keeping all terms) on $\mu$ is shown
for fixed parameters $m_{\tg} = 250 \gev, m_\tq = 1000 \gev$ and for
$a_d = 0.75\, m_\tq$.
We indeed verify from the figure that the impact of the
BaBar  bound~\cite{tmgbound} $BR(\tau \to \mu \gamma)<6.8 \times 10^{-8}$
strongly depends on $\mu$.
For a given $\mu$, the contour where the bound is reached divides the
plane of Fig.~\ref{fig:bb} (left) into an allowed and an excluded
region.
This is illustrated in Fig.~\ref{fig:bbtmg} (right).
We conclude that without specifying $\mu$, large \bbms\ cannot be
excluded by $\tau \to \mu \gamma$:
for instance, for $\mu<-600 \gev$ the impact of
the bound almost disappears.
\footnotetext[3]{A color version of this
plot is included in the slides for this talk, available on the Moriond website.}

\subsection{Bounds from $\Delta B=1$ processes}
While so far there exist only lower bounds on the \bbb mass difference,
one can also try to constrain the CMM model through information on
$b \to s$ transitions such as $B_d \to X_s \gamma$, $B_d \to X_s l^+ l^-$
and $B_d \to \phi K_S$, which have been observed.\cite{gutdelb1,bsgcmm} We have so far not
studied these in comparable detail. However, we can make the
following remarks: $B_d \to X_s \gamma$, like $\tau \to \mu \gamma$,
is sensitive to left-right mixing, which is generally not large
in the CMM model. Furthermore, the chiral flavor structure of the model
would primarily generate the mirror image $Q_{7 \gamma}'$ of the
SM magnetic penguin operator $Q_{7 \gamma}$; then the SM and CMM contributions do
not interfere due to the different chirality structure.\cite{bsgcmm}
In view of the still substantial theoretical uncertainties in this decay,
it is possible that information on this mode poses no additional constraints.
Similar comments apply to $B_d \to X_s l^+ l^-$. Nevertheless, a complete
analysis should include a more detailed treatment.
The branching ratios for loop-induced hadronic decay modes such as
$B_d \to \phi K_S$ suffer from large theoretical uncertainties,
precluding any effective constraint.

\subsection{Mixing-decay interference in $b \to s$ penguin decays}
Unlike the decay rates, the time-dependent CP asymmetries in
penguin-dominated $B_d \to M K_S$ decays with a vector or pseudoscalar meson $M$
in the final state
are theoretically rather clean and close to the one in $B_d \to J/\psi K_S$ in
the Standard Model.
This is due to the fact that, the decay amplitudes being nearly real,
all are dominated by the single weak phase $2 \beta$ entering through
the \bbmd\  amplitude.

While in the latter mode the decay amplitude arises at tree level,
in the former it is loop-induced and both its modulus and weak phase
are sensitive to new physics, destroying the relation.\cite{bphiknp}
Indeed there is a well-known discrepancy between the experimental
world average over penguin modes,
$-\eta_{\rm CP} S_{f} (b \to s\; \mbox{penguin}) = 0.43 \pm 0.07$,
and that over charmonium modes,
$\sin 2 \beta (b \to c \bar c s) = 0.726 \pm 0.037$.\cite{sin2betlit}

While we have not yet performed a detailed numerical analysis,
analytical results obtained within QCD factorization for the relevant
operator matrix elements show that the parity
reflection $Q_{8g}'$ of the SM chromomagnetic penguin operator could
give an $\cO(1)$ correction to the Standard Model amplitude, again with an
unconstrained phase $\theta$ correlated with the \bbms\ phase.
We remark that in the limit where this contribution
dominates the amplitude, the CP asymmetry would measure $\sin (2 \beta
+ 2 \theta)$ in all cases. In the general case of interference
with the Standard Model contribution, there is dependence on the final state.\cite{bspengfs}
A recent generic analysis~\cite{rhpeng} relevant to the CMM model suggests,
however, that right-handed squark mixing can account for the data in
both vector and pseudoscalar final states.

\section{Conclusion}
We have studied a SUSY $SO(10)$ model implying correlations between
neutrino mixing angles and the flavor structure of certain sfermion mass matrices.
Large effects are found possible for \bbms\ and $\tau \to \mu \gamma$. Both are correlated,
but the additional dependence of the latter on the $\mu$-parameter allows to satisfy the
experimental bound while retaining a large mass difference and/or CP violation in the former.
A large chromomagnetic penguin in $b \to s$ penguin decays is suggested within QCD factorization.
Whether this can generate time-dependent CP asymmetries at the observed central values remains
to be resolved through a numerical analysis.

\section*{Acknowledgments}
The author thanks Uli Nierste for a pleasant and inspiring collaboration,
and for his tireless efforts in promoting intercultural exchange and
international affairs. Correspondence with S. Vempati on $BR(\tau\to\mu\gamma)$
is gratefully acknowledged. The author wishes to express his gratitude to the
organizers for an exciting
conference in a pleasant atmosphere, and to the EU Marie Curie Conferences programme
for partial support.
The author's work is supported in part by the DFG SFB/Transregio 9
``Computergest{\"u}tzte Theoretische Teilchenphysik''.

\end{document}